\documentclass[pra,twocolumn,showpacs,amssymb,aps,floatfix]{revtex4}
\newcommand{\be}{\begin{equation}}
\newcommand{\ee}{\end{equation}}
\newcommand{\bea}{\begin{eqnarray}}
\newcommand{\eea}{\end{eqnarray}}
\usepackage{graphicx}
\usepackage{float}
\usepackage{amsmath}
\usepackage{color}
\usepackage{dcolumn}
\usepackage{bm}
\usepackage{hyperref}
\usepackage{braket}
\usepackage{enumerate}

\begin{document}
\title{Real-space probe for lattice quasiholes}

\author{R. O. Umucal\i lar}
\affiliation{%
Department of Physics, Mimar Sinan Fine Arts University, 34380 Sisli, Istanbul, Turkey
}%

\date{\today}

\begin{abstract}
We propose a real-space probe that is based on density measurements to extract distinct signatures of quasihole-like states of bosons experiencing a synthetic magnetic field in a two-dimensional lattice. We numerically show that certain ratios of the mean square radii of the particle cloud, obtainable through the density profile, approach the continuum values expected from Laughlin's ansatz wave functions quickly as the magnetic flux quanta per unit cell of the lattice decrease, even in a small lattice with few particles. This method can equally be used in both ultracold atomic and photonic systems.
\end{abstract}


\maketitle
\section{Introduction}
The interaction of charged particles with a magnetic field lies at the heart of many interesting phenomena in condensed matter physics including the quantum Hall effects \cite{quantum Hall experiment, Yoshioka}. Besides the hallmark conductance quantization, a two-dimensional electron gas in a magnetic field hosts curious physics due to the quasihole and quasiparticle excitations theorized to appear in a fractional quantum Hall (FQH) fluid \cite{Laughlin paper}. The unusual fractional exchange statistics of these excitations, which are called anyons \cite{anyons}, is believed to have a great potential for applications in the field of topological quantum computation \cite{anyon computation review}. 

Advances in quantum simulation with ultracold atomic \cite{cold atoms} and photonic systems \cite{light review} have encouraged researchers to look for the FQH physics in analog systems, which provide a more controllable environment than their electronic counterparts. Integration of strong inter-particle interactions with the recently created synthetic magnetism for neutral particles \cite{Hofstadter experiment 1, Hofstadter experiment 2, synthetic field atoms, strain induced LLs, Hafezi edge experiment, Simon Landau levels, topological photonics} seems within reach of current experimental capabilities \cite{Simon microwave, Roushan, Greiner FQH}, paving the way to realize the FQH physics with atoms and photons soon. 

The possibility of forming periodic structures like optical lattices in ultracold atomic systems and coupled cavity arrays in photonic ones has a constructive effect on the experimental realization of FQH physics. Most importantly, inter-particle interactions can be greatly enhanced due to the confinement of particles in lattice sites, thereby increasing the energy gap that protects the ground state from external perturbations. Motivated by this advantage and the promising methods for creating synthetic magnetic fields, the lattice version of the FQH effect has been vigorously investigated for both ultracold atomic \cite{Lukin FQH, optical lattice FQH, Hafezi torus} and photonic systems \cite{photonic lattice FQH}. As a parallel development, fractional Chern insulators (FCIs) with topological flat bands, a broader class of systems which do not require a uniform magnetic field for FQH-like effects to appear in lattices, have become a subject of intense study in recent years \cite{FQH flat band}. Exchange properties of quasihole excitations have also been examined for both FCIs \cite{Regnault} and other lattice FQH models constructed to have the quasihole state as the system's ground state \cite{lattice quasiholes}.

So far, several experimental methods involving density measurements have been put forward for the identification of FQH-like states both in continuum and in lattices, including the observation of a flat density profile suggesting incompressibility, quasiholes with an estimated size \cite{Regnault}, and fractional density depletion at the quasihole position \cite{Nur}. Also, we have recently proposed a real-space method for observing the anyonic statistics of quasiholes in a system of trapped particles in continuum \cite{time-of-flight paper}.

In this work, we show how by determining the mean square radii of various many-particle states through a density measurement in the lattice one can infer, in an unambiguous manner, whether a small number of interacting particles exist in a quasihole-like state. By numerically studying the repulsive Hofstadter-Hubbard model \cite{Lukin FQH} in the presence of an impurity potential, we found that certain ratios of the mean square radii of the particle cloud quickly approach the continuum expectations as the magnetic flux quanta per unit cell decrease, even in a small lattice. We argue that the dependence of a global observable like the mean square radius on the number of particles, in a measurably distinct way for small systems, can provide useful supplementary information about the underlying microscopic physics in addition to other local signatures such as the quasihole size and fractional density depletion. Moreover, by explicitly showing the agreement between the continuum expectations and lattice results for small systems, we provide a reasonable conjecture that mean-square-radii measurements, originally proposed for a continuous system \cite{time-of-flight paper}, can also be utilized to observe quasihole anyonic statistics in moderate-sized lattices.

In order to avoid edge effects for the finite system that we study and to focus on the bulk properties, we use periodic boundary conditions in our numerical simulations. Such boundary conditions for two-dimensional lattices might be realized in cold-atom systems by creating a torus surface using spatially shaped laser beams \cite{Hafezi torus} and in photonic systems by connecting the opposite edges of the finite system possibly with wave guides. From an experimental point of view, however, it is easier to impose an overall trapping potential on a finite lattice, which confines the particles in the center of the system, than to implement periodic boundary conditions. Provided that the number of magnetic flux quanta per particle in a large enough region away from the system edge is the correct bulk value, we believe the mean-square-radius approach should still work in an appropriate limit without being hindered by the discrete nature and moderate size of the lattice. We defer the study of this case to a future work. The density measurements we rely on can be straightforwardly performed in cold-atom systems via time-of-flight methods \cite{time of flight measurement} or by using quantum gas microscopes with single-site resolution, which are particularly suitable for two-dimensional optical lattices \cite{quantum gas microscope}, and in the photonic context via standard imaging techniques that collect scattered light from individual cavities \cite{strain induced LLs, Hafezi edge experiment}.

\section{The Model}
We start with the noninteracting Hamiltonian for particles hopping in a square lattice perpendicularly pierced by a uniform synthetic magnetic field along the $z$ direction:
\bea 
H_0 = - t\sum_{\langle ij\rangle}\left(e^{i2\pi \phi_{ij}} 
c^\dag_{i}c_{j}+\text{h.c.}\right), 
\label{eq:Hofstadter Hamiltonian}
\eea
where $c^{\dagger}_i$ ($c_j$) creates (annihilates) a 
boson at site $i$ ($j$), \text{h.c.} is the Hermitian conjugate, and $t > 0$ is the hopping amplitude between nearest-neighbor sites $\langle ij\rangle$ with coordinates ${\bf r}_i$ and ${\bf r}_j$. The hopping phase is given by $\phi_{ij} = (1/\phi_0) \int_{{\bf r}_j}^{{\bf r}_i} {\bf A}\cdot d{\bf r}$, where integration path is a straight line, $\phi_0 = h/q_0$ is the magnetic flux quantum for a synthetic charge $q_0$ and ${\bf A} = -By{\bf \hat{x}}$ is the Landau gauge vector potential corresponding to an effective magnetic field strength of $B$. The quantities $q_0$ and $B$ are merely introduced to make the synthetic-real analogy complete; the experimentally relevant quantity is the phase $\phi_{ij}$ itself. We also define the magnetic flux quantum per unit cell of the lattice as $\phi = Ba^2/\phi_0$, where $a$ is the lattice constant. In this model, the wave function of a particle traversing a loop around the unit cell acquires the Aharonov-Bohm phase factor $\exp(i2\pi\phi)$. When $\phi = p/q$, with $p$ and $q$ being relatively prime integers, the single-particle energy band in the absence of a magnetic field is split into $q$ sub-bands yielding the fractal Hofstadter butterfly spectrum \cite{Hofstadter}.  

We consider repulsive on-site interactions between particles, modeled by the interaction Hamiltonian $ H_I = (U/2) \sum_i n_i(n_i-1)$, where $n_i = c^{\dagger}_i c_i$ is the number operator and $U>0$. The overall Hamiltonian is therefore given by $H = H_0+H_I$, which is simply the Bose-Hubbard Hamiltonian \cite{Bose-Hubbard} including the effect of the synthetic magnetic field through complex hopping amplitudes. This Hamiltonian has been investigated in numerous works \cite{Lukin FQH, optical lattice FQH, Hafezi torus} and its ground state has been found to have a very large overlap with the Laughlin state (generalized for torus boundary conditions; cf. Appendix \ref{LWF Torus}) for the appropriate filling fraction $\nu = N/N_{\phi}$ in the so-called continuum limit $\phi \ll 1$. Here, $N$ is the number of particles and $N_{\phi}$ is the number of flux quanta contained in the lattice. For a filling fraction $\nu = 1/m$, where $m$ is an even integer in the case of bosons, the ground state turns out to be $m$-fold degenerate for torus boundary conditions \cite{FQH torus}. We will focus on the simplest $\nu = 1/2$ case in the following discussion, as long-range interactions might be necessary to separate the degenerate ground states from the excited ones for smaller filling fractions \cite{Lukin FQH}.

Although we perform exact diagonalization of small systems for benchmarking purposes, we use a projection method \cite{Regnault} in momentum (${\bf k}$) space to deal with larger systems for which exact diagonalization is time consuming if not totally out of reach. For this purpose, we first solve the single-particle problem in ${\bf k}$-space. We define the Fourier transform of $c_i$ in an $L_x \times L_y$ lattice (lattice constant $a$ set to unity)
\bea 
c_{{\bf k}\beta} = \sqrt{\frac{q}{L_x L_y}} 
\sum\limits_{ s = 0}^{\frac{L_y}{q}-1}\sum\limits_{ i_x = 0}^{L_x-1}
c_{s\beta i_x}e^{-ik_y s q}e^{-i k_x i_x},
\label{eq:k-space c}
\eea
where $c_{s\beta i_x} \equiv c_i$, the $y$-coordinate of the $i$th site is given by $i_y = sq+\beta$ with $s = 0,\ldots,L_y/q-1$ labeling a magnetic unit cell that covers $q$ sites along the $y$-direction, and $\beta = 0,\ldots, q-1$ is the index of a site inside the unit cell. With this choice of the unit cell, a $q$-site translation along the $y$-direction gives a total hopping phase factor of unity (which is equivalent to the zero-flux case) and the Brillouin zone is reduced to $k_x \in [0,2\pi)$ and $k_y \in [0,2\pi/q)$. By also imposing periodic boundary conditions (PBCs), the noninteracting Hamiltonian is written as $H_0 = \sum_{{\bf k} \alpha \beta} c^{\dagger}_{{\bf k}\alpha }M_{\alpha \beta}({\bf k})c_{{\bf k}\beta}$, where $M({\bf k})$ is a $q\times q$ matrix with components $M_{\alpha \alpha}({\bf k}) = -2t\cos(2\pi\alpha\phi+k_x)$, $M_{\alpha (\alpha \pm 1)}({\bf k}) = -t$, $M_{1 q}({\bf k}) = M_{q 1}^{\ast}({\bf k}) =  -t\exp(-ik_yq)$ and all the remaining matrix elements are zero. After diagonalizing $M({\bf k})$ we get the single-particle energies $\epsilon_n({{\bf k})}$, which yield the Hofstadter butterfly when plotted as a function of $\phi$, and the corresponding eigenvectors $g^{(n)}({\bf k})$, where $n = 1,2,\ldots,q$ is the band index. The Hamiltonian $H$ can now be written in terms of the operators $d_{{\bf k}n} = \sum_{\beta}g^{(n)\ast}_\beta({\bf k})c_{{\bf k}\beta}$ that diagonalize $H_0$ as
\begin{multline}
H = \sum_{{\bf k}n}\epsilon_n({\bf k})d^{\dagger}_{{\bf k}n} d_{{\bf k}n} 
\\ + \frac{Uq}{2 L_x L_y}\!\!\!\!\sum_{\substack{{\bf k}{\bf k}^{\prime}{\bf Q}\\n_1n_2n_3n_4}}\!\!\!\!d^{\dagger}_{{\bf k}+{\bf Q},n_1}d^{\dagger}_{{\bf k}^{\prime}-{\bf Q},n_2}d_{{\bf k}^{\prime}n_3}d_{{\bf k}n_4} \\ \times g^{(n_1)\ast}_\beta({\bf k}+{\bf Q})g^{(n_2)\ast}_\beta({\bf k}^{\prime}-{\bf Q})g^{(n_3)}_\beta({\bf k}^{\prime})g^{(n_4)}_\beta({\bf k}),
\label{eq:k-space Hamiltonian}
\end{multline}
where $g^{(n)}_\beta({\bf k})$ stands for the $\beta$th component of $g^{(n)}({\bf k})$.

In order to lessen the computational burden, we choose to describe the physics in the lowest band of the single-particle spectrum, by keeping only $n = 1$ terms in the Hamiltonian (\ref{eq:k-space Hamiltonian}). For this projection to be valid, we require that the strength $U$ of the inter-particle interactions be small enough to avoid scattering of particles to higher bands \cite{note 2}. Note that this approximation is similar to the lowest Landau level (LLL) approximation in continuum, where the interaction-induced gap is much smaller than the separation between Landau levels. In the mean time, $U$ should not be too small as interactions are necessary to observe Laughlin-type strongly-correlated ground states. We also add to the Hamiltonian $H$ a simple repulsive impurity potential $V_{\rm imp} = Vn_i$ with a sufficiently large strength $V>0$ to pin a quasihole on the $i$th site. The ${\bf k}$-space form of $n_i$ should also be projected to the lowest band.

\section{Mean-square-radius Approach}
In this section, we lay out our approach to find the signatures of Laughlin-type correlations through a mean-square-radius measurement in the lattice. First, we briefly overview the situation in continuum. 

In order to provide a microscopic explanation for the FQH effect for the two-dimensional electron gas at filling fraction $\nu = 1/3$, Laughlin put forward the following ansatz wave function composed of single-particle LLL wave functions \cite{Laughlin paper}
\bea
\Psi_{\rm FQH}(\zeta_1, \ldots, \zeta_N) \propto \prod_{j<k}(\zeta_j-\zeta_k)^m e^{-\sum_{i = 1}^N|\zeta_i|^2/4\ell_B^2},\label{WF_FQH}
\eea
where $N$ is the number of particles in the system, $\zeta_j = x_j+iy_j$ is the complex-valued coordinate of the $j$th particle, $m = 1/\nu = 3$, and $\ell_B = \sqrt{\hbar/eB}$ is the magnetic length. This ansatz readily extends to other filling fractions $\nu = 1/m$, $m$ being an odd (even) integer for fermions (bosons), yielding the correct symmetry for the wave functions. Indeed, the bosonic extension of the FQH physics has been successfully carried out to explore the ground states of rotating atomic condensates \cite{Bosonic FQH}. There is also a simple ansatz for the quasihole wave function as follows \cite{Laughlin paper}
\bea
\Psi_{\rm qh}(\{\zeta_i\},\mathcal{R}) \propto  \prod_{i=1}^N(\zeta_i\!-\!\mathcal{R})\Psi_{\rm FQH}(\zeta_1, \ldots, \zeta_N) \label{1QH wf},
\eea
where $\mathcal{R}$ is the complex-valued coordinate of the quasihole that could be pinned by impurities in electronic systems or repulsive localized potentials in ultracold atomic systems. Equations (\ref{WF_FQH}) and (\ref{1QH wf}) were numerically verified to accurately describe the low-energy physics of the relevant systems.

In our recent work \cite{time-of-flight paper}, we proposed to observe quasihole anyonic statistics by measuring the mean square radius $\langle r^2\rangle = \int r^2 n({\bf r})d^2{\bf r}/N$, where $n({\bf r})$ is the particle density. For a many-particle wave function described in the LLL, it is possible to relate $\langle r^2\rangle$ to the mean total angular momentum $\langle L_z \rangle$ along the $z$ axis through the following relation \cite{Ho Mueller}
\bea
\langle r^2\rangle = \frac{2\ell_B^2}{N}\bigg(\frac{\langle L_z\rangle}{\hbar} + N\bigg).
\label{r2 Lz}
\eea
It is this relation that makes a real-space observation of the exchange statistics possible as the Berry phase \cite{Berry phase} of particle braiding is given by $2\pi\langle L_z \rangle$ (defined modulo $2\pi$) \cite{time-of-flight paper}.

We now investigate whether we can exploit Eq. (\ref{r2 Lz}) to predict the correlated nature of the ground state in the lattice. It is by no means clear from the outset that an equation valid in an infinite continuous space could be used to describe a discrete system on a torus. However, it is plausible to conjecture that in a limit where the discreteness and boundary effects are not much pronounced, such an equation can provide approximate but still useful information. There is ample analytic and numerical evidence that the ground state wave functions on a torus are the appropriately generalized versions of those in Eqs. (\ref{WF_FQH}) and (\ref{1QH wf}) for PBCs \cite{FQH torus}. In addition, it is known that lattice ground states can be constructed with high fidelity by a discrete sampling of the continuum wave functions at lattice points as long as $\phi \ll 1$ \cite{Lukin FQH}; that is, when the cyclotron orbit characterized by $\ell_B$ encircles a sufficiently large number of unit cells with side length $a$, as can be seen through the relation $(\ell_B/a)^2 = 1/2\pi\phi$.

We first identify what Eq. (\ref{r2 Lz}) means for the continuum wave functions. The Laughlin (L) state and the quasihole (QH) state with the quasihole pinned at the origin ($\mathcal{R} = 0$) are both total angular momentum eigenstates with eigenvalues $N(N-1)\hbar$ and $[N(N-1)+N]\hbar = N^2\hbar$, respectively, for the case of $m = 2$. Using these values we arrive at
\bea
\langle r^2\rangle_{\rm L}^{N,\phi} = 2N\ell_B^2 = \frac{N}{\pi\phi}a^2,\label{r^2 L}\\
\langle r^2\rangle_{\rm QH}^{N,\phi} = 2(N+1)\ell_B^2 = \frac{N+1}{\pi\phi}a^2.
\label{r^2 QH}
\eea
If the system can be brought sufficiently close to its lowest energy configuration, these $\langle r^2\rangle$ values will be peculiar to Laughlin and quasihole states as all other states with same angular momenta lie above a sizable energy gap. In Eqs. (\ref{r^2 L}) and (\ref{r^2 QH}) we also introduced the lattice constant $a$ to establish a link between continuum and lattice physics, bringing the flux quantum per plaquette $\phi$ to the scene. One may argue that in the limit $\phi \ll 1$, the lattice values of $\langle r^2\rangle$ would approach the continuum expectations given in Eqs. (\ref{r^2 L}) and (\ref{r^2 QH}). However, the artificiality of boundary conditions, which is more pronounced for small systems, prevents this expectation from being truly realized. In order to alleviate this obstacle, we propose to look at certain ratios of $\langle r^2\rangle$ for different states, conjecturing that these ratios could be less sensitive to the boundary conditions. As will be seen in the next section, this is indeed the case. Incidentally, looking at ratios could also be experimentally more viable, as they are more robust against fluctuations. Specifically, we define
\bea
R_{\rm L/QH} \equiv \langle r^2\rangle_{\rm L}^{N,\phi}/\langle r^2\rangle_{\rm QH}^{N,\phi} = N/(N+1),\label{R L/QH}\\
R_{\rm QH/QH} \equiv \langle r^2\rangle_{\rm QH}^{N,\phi}/\langle r^2\rangle_{\rm QH}^{N+1,\phi^{\prime}} = \frac{(N+1)\phi^{\prime}}{(N+2)\phi}.\label{R QH/QH}
\eea
While the ratio in Eq. (\ref{R L/QH}) compares $\langle r^2\rangle$ values for the Laughlin and quasihole states with same $N$ and $\phi$, the one in Eq. (\ref{R QH/QH}) is for two quasihole states differing by one particle and experiencing different fluxes; namely, $\phi$ and $\phi^{\prime}$. The last equalities in Eqs. (\ref{R L/QH}) and (\ref{R QH/QH}) follow from the continuum expectations given in Eqs. (\ref{r^2 L}) and (\ref{r^2 QH}). In the next section, we compare the continuum expectations with the numerical results for the lattice. We also discuss in detail how we choose the value of the flux and the lattice size to obtain Laughlin and quasihole states.

\section{Numerical Results}
In an $L_x\times L_y$ lattice, the total number of flux quanta is $N_{\phi} = L_xL_y\phi$. We consider simple fractions $\phi = 1/q$ and choose $L_y = q$ so that only one magnetic unit cell fits along the $y$-axis. Therefore, $N_{\phi}$ equals $L_x$ in our model. For simplicity and in order to deal with a symmetric system, we also choose $L_x = q$ for the Laughlin state, yielding $N_{\phi} = q$ and $N = N_{\phi}/2 = q/2$. We found from the exact diagonalization of the $N = 2,3$ systems in real-space that the ground state is twofold degenerate and the overlap between any of the two degenerate ground states and an optimal linear combination of the two Laughlin states generalized for PBCs is $\sim 99\%$ for a sufficiently large $U$.

When it comes to creating the quasihole state, in addition to applying the impurity potential $V_{\rm imp}$ to remove one half of a particle at the position of the quasihole (cf. Appendix \ref{QH pinning}), we must enlarge the system to the extent that it exactly contains one more flux quantum; that is, the new number of flux quanta becomes $N_{\phi}^\prime = N_{\phi}+1 = 2N+1$. We do this by increasing $L_x$ by one, thereby introducing $q$ sites along the $y$-axis, which brings an additional $q\phi = 1$ flux quantum to the lattice as required \cite{note}. 

In order to calculate $\langle r^2\rangle = \sum_i r_i^2\langle n_i\rangle/N$ one first needs to find the distance $r_i = a\sqrt{i_x^2+i_y^2}$ of each lattice point from a specified origin by paying attention to PBCs (cf. Appendix \ref{Measuring r2 on a Torus}). In the presence of a quasihole pinning potential, we take the origin to be the site at which the pinning potential is localized; when there is no such potential as in the case of the Laughlin state, any site can be chosen as the origin without altering the results we present in Table \ref{Table1}. Since the ground state manifold is twofold degenerate for a sufficiently large $U$, the expected value $\langle n_i\rangle$ is averaged over these two states.

\begin{table}[h!]\footnotesize
\centering
\begin{tabular}{ |c||c|c||c|c||c|c||c|c|}
\hline
&\multicolumn{2}{c||}{$\langle r^2\rangle^{N,\phi}_{\rm L}/a^2$} &\multicolumn{2}{c||}{$\langle r^2\rangle^{N,\phi}_{\rm QH}/a^2$} & \multicolumn{2}{c||}{$R_{\rm L/QH}$} &\multicolumn{2}{c|}{$R_{\rm QH/QH}$} \\
 \hline
 &{\rm cont.}&{\rm lat.}&{\rm cont.}&{\rm lat.}&{\rm cont.}&{\rm lat.}&{\rm cont.}&{\rm lat.}\\
 \hline
 N=2&2.547&3.000&3.820&\begin{tabular}{@{}c@{}}3.571 \\ (4.050)\end{tabular}&0.667&\begin{tabular}{@{}c@{}}0.840 \\ (0.741)\end{tabular}&0.500&\begin{tabular}{@{}c@{}}0.437 \\ (0.493)\end{tabular}\\
 \hline
 N=3&5.730&6.333&7.639&\begin{tabular}{@{}c@{}}8.171 \\ (8.210)\end{tabular}&0.750&\begin{tabular}{@{}c@{}}0.775 \\ (0.771)\end{tabular}&0.600&\begin{tabular}{@{}c@{}}0.603 \\ (0.606)\end{tabular}\\
 \hline
 N=4&10.19&11.00&12.73&13.54&0.800&0.812&0.667&0.670\\
 \hline
 N=5&15.92&17.00&19.10&20.21&0.833&0.841&0.714&0.717\\
 \hline
 N=6&22.92&24.33&26.74&28.20&0.857&0.863&0.750&0.752\\
 \hline
 N=7&31.19&33.00&35.65&37.52&0.875&0.879&-&-\\
 \hline
\end{tabular}
\caption{Continuum expectations (cont.) calculated using Eqs. (\ref{r^2 L}) to (\ref{R QH/QH}) and the corresponding numerical results for the lattice (lat.) with $U = t$, $V = 5t$ (0) in the QH (L) case. Results for $\langle r^2\rangle^{N,\phi}_{\rm QH}$ given in parentheses are the converged values obtained via exact diagonalization using $U = 50t$.}
\label{Table1}
\end{table}

The results for $N = 2,3$ come from exact diagonalization and the rest are found with the lowest-band approximation. We observed that $\langle r^2\rangle$ converges very quickly to the quoted values for an interaction strength $U<t$ for all $N$ but $N=2$. For large enough $U$, the two lowest-energy states of the $N=2$ system in the presence of the impurity potential are only nearly degenerate. Combined with the smallness of the system, this leads to a noticeably different $\langle r^2\rangle$ for these states even when the results converge for large $U$ (cf. Appendix \ref{r2  and U}).

\begin{figure}[ht]
\includegraphics[width=0.48\textwidth]{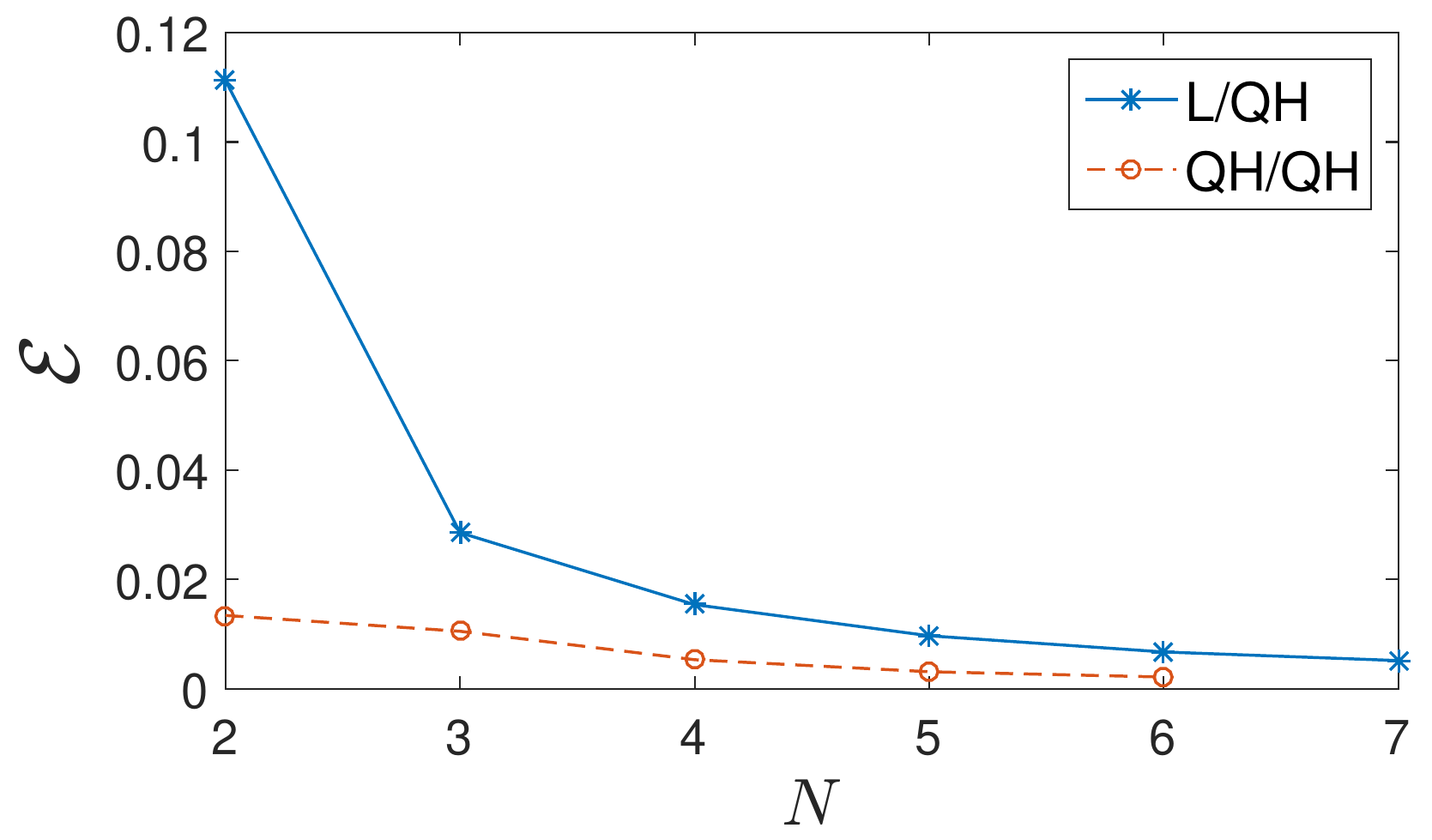}
\caption{Relative error $\mathcal{E}$ between lattice ratios and the corresponding continuum expectations as a function of the particle number $N$. We used the $U = 50t$ results for $N = 2,3$. Blue stars are for $R_{\rm L/QH}$ and red circles for $R_{\rm QH/QH}$. Lines are drawn as a guide.}
\label{Fig1}
\end{figure}

As can be noticed from Table \ref{Table1}, $\langle r^2\rangle$ for the Laughlin case takes some integer and rational values. This is simply because, as numerically verified, the site densities $\langle n_i\rangle$ averaged over two degenerate Laughlin states are very nearly the same ($\langle n_i \rangle \simeq N/L_xL_y$) just as the uniform bulk of the continuum version and as a result $\langle r^2\rangle^{N,\phi}_{\rm L}/a^2$ is given by the sum $\sum_i r_i^2 /L_xL_ya^2=(2N^2+1)/3$, with $L_x=L_y=q=2N$. Still, since the filling fraction is fixed and most of the contribution to $\langle r^2 \rangle$ comes from the uniform bulk, continuum results given by $N/\pi\phi = 2N^2/\pi$ can be considered close to the lattice ones, given the discrete nature of the lattice. Lattice and continuum results for $\langle r^2\rangle^{N,\phi}_{\rm QH}$ are also comparable. More interesting, however, are the results for  $R_{\rm L/QH}$ and $R_{\rm QH/QH}$, for which the continuum expectations in Eqs. (\ref{R L/QH}) and (\ref{R QH/QH}) yield $N/(N+1)$ and $N/(N+2)$, respectively, for the parameters at hand [$\phi = 1/2N$, $\phi^{\prime}=1/2(N+1)$]. The agreement between results for $R_{\rm QH/QH}$ is especially remarkable. In Fig. (\ref{Fig1}), we plot the relative error $ \mathcal{E} \equiv |R_{\rm lat.}/R_{\rm cont.}-1|$ between lattice ratios $R_{\rm lat.}$ and the corresponding continuum expectations $R_{\rm cont.}$ in order to see better the convergence of results as the continuum limit $\phi \sim 1/2N \ll 1$ is approached. While the case for $N = 2$ can be considered anomalous as discussed above, $\mathcal{E}$ quickly gets smaller to reach the value $\sim 0.5\%$ for the ratio $R_{\rm L/QH}$ with $N = 7$ and $\sim 0.2\%$ for $R_{\rm QH/QH}$ evaluated for the quasihole states with $N = 6,7$.

We believe that the good agreement observed for certain $\langle r^2\rangle$ ratios results from satisfying several gross features of the quasihole state. It seems that as long as the filling fraction $N/N_{\phi}$ is the correct one so as to remove a fraction (here one half) of the particle from the quasihole position and increase the system area accordingly, and the density around the quasihole has a sufficient radial symmetry (although discrete) for $\langle r^2 \rangle$ to be a meaningful quantity, the effect of the underlying lattice on the ratios we investigated quickly diminishes as the continuum limit $\phi \ll 1$ is approached. The relation $R_{\rm L/QH} = N/(N+1)$ can also be shown to follow from a simple disk model for the density in continuum, emphasizing that a detailed knowledge of the actual density profile of the Laughlin and quasiholes states is not required when it comes to calculating this specific ratio (cf. Appendix \ref{Disk model}).

\section{Conclusion}
We proposed a method that depends on real-space density measurements for obtaining clear signatures of quasihole states of lattice bosons in a synthetic magnetic field. We provided strong numerical evidence that certain ratios of the mean square radii of Laughlin- and quasihole-like states in the lattice approach the values expected from continuum physics, even in a small system, when the discrete nature of the lattice becomes less discernible as in the so-called continuum limit characterized by small magnetic flux quanta per unit cell. We believe our proposal will be especially useful to identify quasihole states in the first experimental realizations, which will most probably involve a small number of particles and lattice sites.

The agreement we found between the continuum expectations and lattice results also encourages us to anticipate that mean-square-radii measurements in a moderate-sized lattice can still be used as a means to observe quasihole anyonic statistics. Therefore, as a future direction, we plan to investigate the statistical phase in larger lattices, which allow for larger separation between quasiholes, by employing Monte Carlo methods. Another interesting venue could be the search of similar signatures in various fractional Chern insulators.

\section*{ACKNOWLEDGMENTS}
This work was supported by the BAGEP Award of the Science Academy (Turkey). The author warmly acknowledges useful discussions with I. Carusotto, N. Ghazanfari, E. Macaluso, and M. Oktel.

\appendix

\section{Laughlin Wave Function on a Torus}
\label{LWF Torus}

In an $L\times L$ torus geometry, the $\nu=1/2$ Laughlin wave function of $N$ particles in the Landau gauge $\mathbf{A} = -By\hat{\bf x}$ has the form \cite{FQH torus}
\bea
\label{Laughlin}
\Psi^{(l)}(\zeta_1,...,\zeta_N) &=& \mathcal{N}_L F^{(l)}_{\rm CM}(Z)e^{-\pi\alpha \sum_j y_j^2} \nonumber \\
&\times& \prod_{j<k}^{N}\left(\vartheta \left[
\begin{array}{c}
\frac{1}{2} \\
\frac{1}{2} \\
\end{array}
\right]
\Big(\frac{\zeta_j-\zeta_k}{L}\Big{|}i\Big)\right)^2,
\eea
where $\zeta_k = x_k+i y_k$ is the complex coordinate of the $k$th particle in units of lattice spacing $a$, $Z = \sum_{j}\zeta_j$, $\phi = Ba^2/(h/e)$ is the magnetic flux quanta per plaquette and $\mathcal{N}_L$ is the normalization factor. The part containing relative coordinates is written in terms of the elliptic $\vartheta$ (theta) functions $\vartheta [\substack{c\\d}] (\zeta|\tau) = \sum_n e^{i\pi\tau(n+c)^2+2\pi i(n+c)(\zeta+d)}$, where $n$ runs over all integers. The center-of-mass part is also given by elliptic functions:
\be
F^{(l)}_{\rm CM}(Z) = \vartheta \left[
\begin{array}{c}
l/2+(N_{\phi}-2)/4\\
-(N_{\phi}-2)/2\\
\end{array}
\right]
\left(\frac{2Z}{L}\Big{|}2i\right).
\ee
Here, $N_{\phi}$ is the number of flux quanta contained in the $L\times L$ lattice and the label $l =0,1$ indicates the two degenerate ground states at filling $\nu=N/N_{\phi}=1/2$.

\section{Quasihole Pinning}
\label{QH pinning}

\begin{figure}[htbp]
\includegraphics[width=0.48\textwidth]{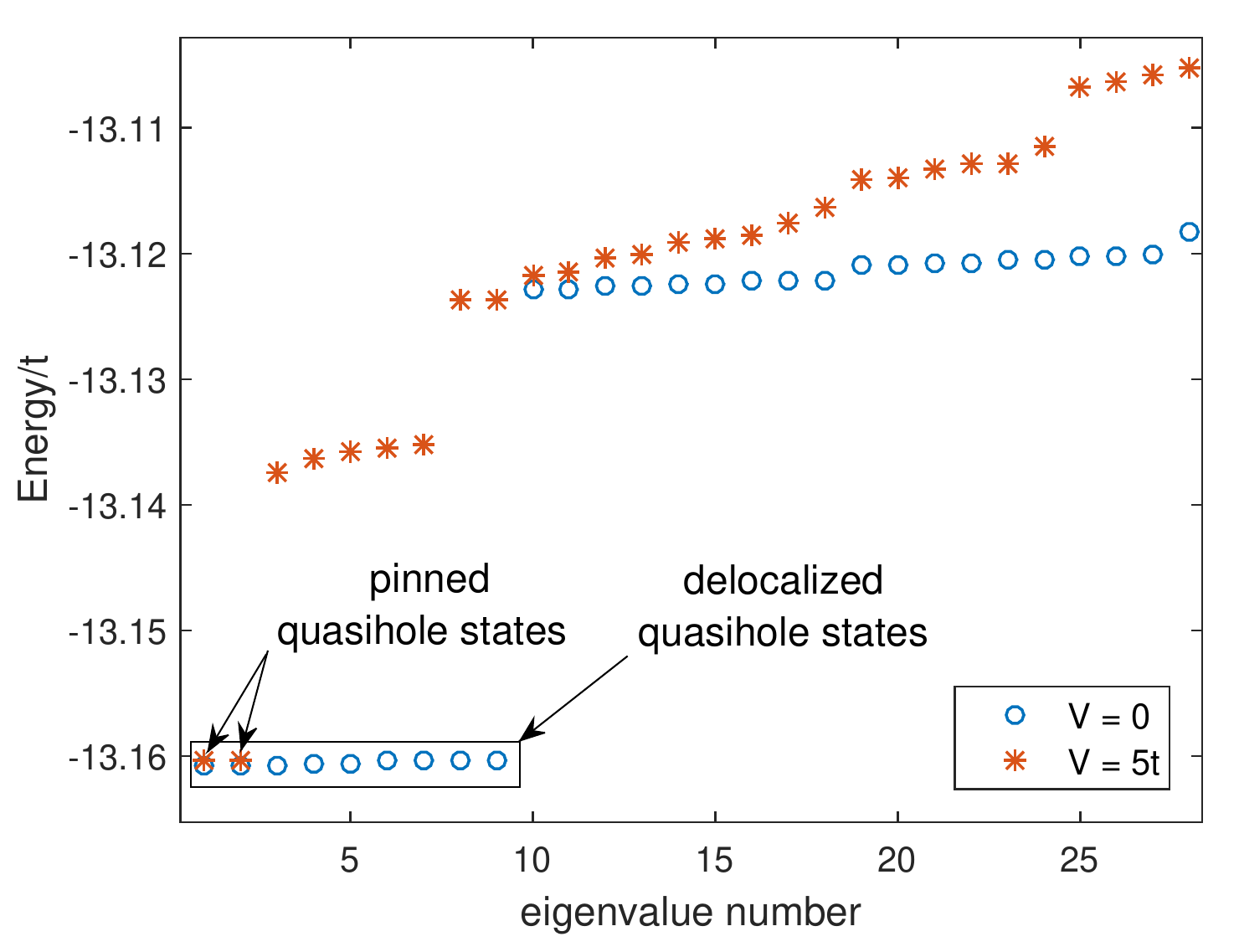}
\caption{Energy spectra for $N = 4$ particles in a lattice with $L_x = 9$, $L_y = 8$, and $\phi = 1/8$. Eigenvalues are ordered from the smallest to the largest. Periodic boundary conditions are imposed. Strength of interactions between particles is $U = t$.  In the absence of an impurity potential ($V=0$), there is a nearly degenerate manifold of delocalized quasihole states. The number of such states is determined by $L_x$. When an impurity potential is set in ($V = 5t$), the quasihole in two of these states is pinned at the position of the localized impurity potential, their energy being almost unchanged.}
\label{SuppFig1}
\end{figure}

\begin{figure}[htbp]
\includegraphics[width=0.48\textwidth]{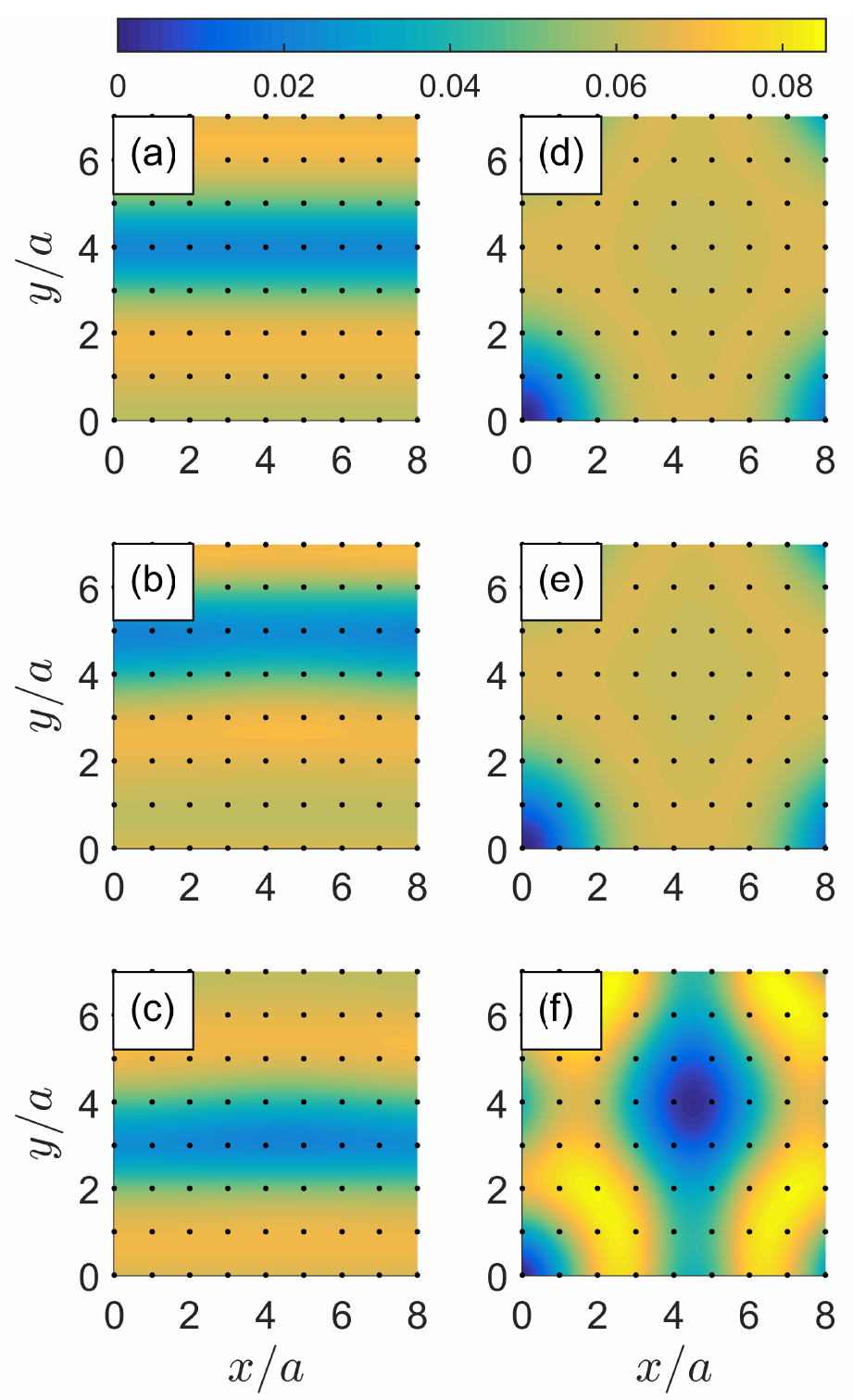}
\caption{Density profiles $\langle n_i\rangle$ (interpolated for better visualization) for the system described in Fig. \ref{SuppFig1}. Panels (a)--(c) are for the first three nearly degenerate delocalized quasihole states in the absence of an impurity potential ($V=0$). Panels (d)--(f) show the density profile for the first three eigenstates when there is an impurity potential with strength $V = 5t$. While the first two profiles in (d) and (e) correspond to the degenerate quasihole states where the quasihole is pinned at the origin, the third one in (f) corresponds to an excited state.}
\label{SuppFig2}
\end{figure}

In a lattice with the correct filling fraction to obtain a quasihole and in the presence of interactions, there appears a degenerate manifold of delocalized quasihole states if there is no impurity potential \cite{Regnault}. When the impurity potential is turned on, the quasihole in two of these delocalized quasihole states gets pinned at the position of the impurity potential localized on a specific site, without any appreciable energy cost (see Fig. \ref{SuppFig1}). This twofold degeneracy is the same one observed for the Laughlin states generalized for torus boundary conditions. The energies of the rest of the delocalized quasihole states are raised and the ground-state manifold becomes isolated. In Fig. \ref{SuppFig2}, we plot the density profiles before and after the impurity potential is turned on, showing the pinning of the quasiholes. When there is no impurity potential, the quasiholes seem to be delocalized along the direction of the greater side of the rectangle, forming stripes. We checked that this is not always the case when the system is kept symmetric with $L_x = L_y = q+1$ and $\phi = 1/(q+1)$.

\section{Measuring $\langle r^2\rangle$ on a Torus}
\label{Measuring r2 on a Torus}

\begin{figure}[htbp]
\includegraphics[width=0.5\textwidth]{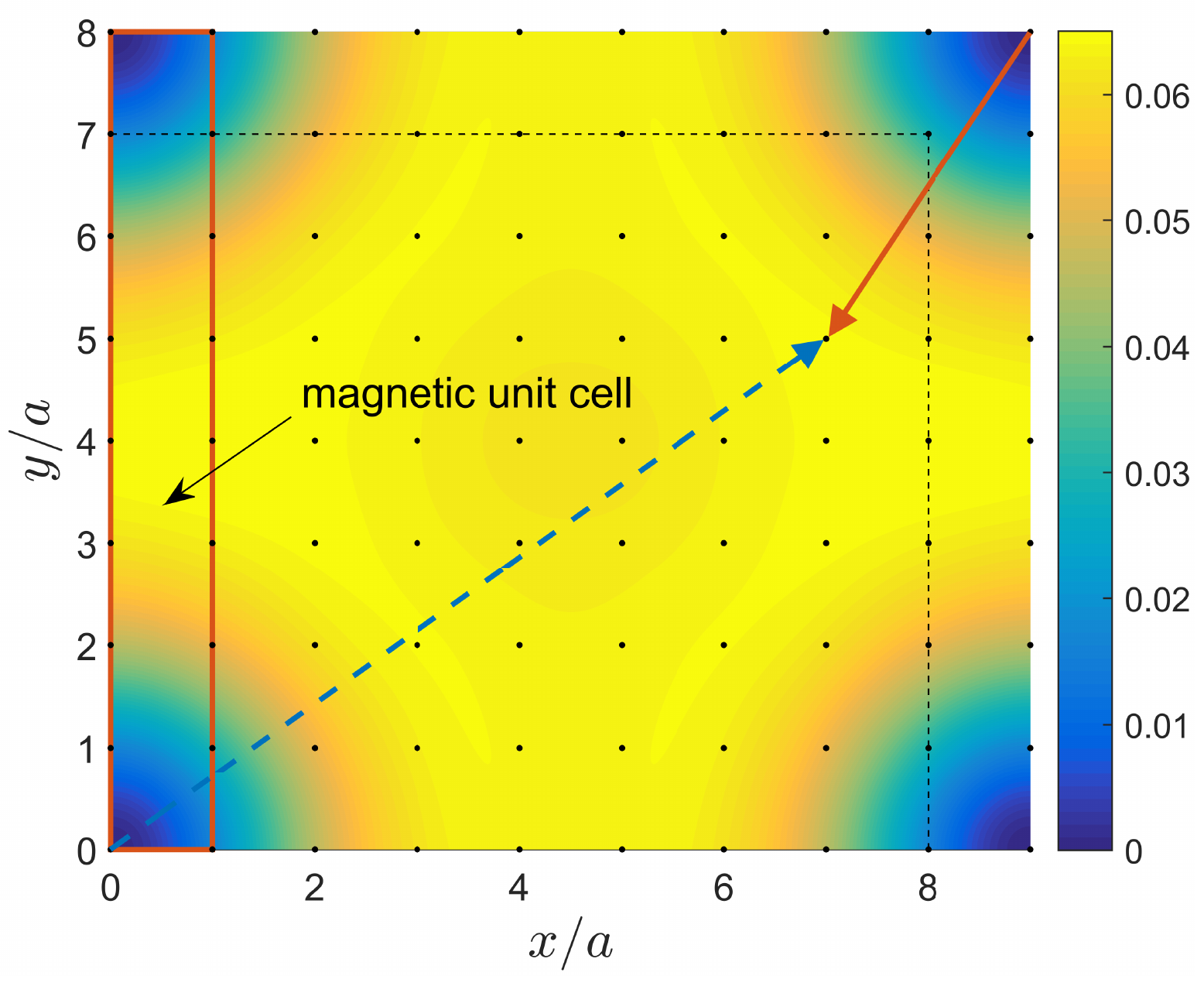}
\caption{Density profile $\langle n_i\rangle$ for the quasihole state (interpolated for better visualization) with $N = 4$, $L_x = 9$, $L_y = 8$, and $\phi = 1/8$. Interaction strength is $U = t$. Impurity potential with strength $V = 5t$ is localized at (0,0). The $1\times 8$ magnetic unit cell is shown by the red rectangle. The distance from the origin of point (7,5) should be taken as the magnitude of the red solid vector and not that of the blue dashed one due to periodic boundary conditions (see the text).}
\label{SuppFig3}
\end{figure}

Here, we briefly explain how we calculate $\langle r^2\rangle$ in a lattice with periodic boundary conditions. In Fig. \ref{SuppFig3}, we show as a generic example the (interpolated) density profile $\langle n_i\rangle$ for the $N=4$ system described in Fig. \ref{SuppFig1} with the impurity potential localized at the origin $(0,0)$. The density profile has a nice radial symmetry around the impurity although the lattice itself is slightly asymmetric. Although we did not make an explicit overlap calculation, the density profile and the twofold degenerate ground-state manifold isolated from excited states are strong indications that we have the quasihole state. In order to calculate $\langle r^2\rangle = \sum_i r_i^2\langle n_i\rangle/N$ we need to find the distance $r_i = a\sqrt{i_x^2+i_y^2}$ of each lattice point from the origin that lies inside the square delineated by dashed lines in Fig. \ref{SuppFig3} defining our finite system. As the four corner points (0,0), (9,0), (0,8), and (9,8) can equivalently be taken as the origin for periodic boundary conditions, the logical choice to define the distance $r_i$ is to take it as the magnitude of the shortest vector connecting the $i$th lattice site with these four points.

\section{Dependence of $\langle r^2\rangle_{\rm QH}$ on the interaction strength}
\label{r2  and U}

\begin{figure}[htbp]
\includegraphics[width=0.48\textwidth]{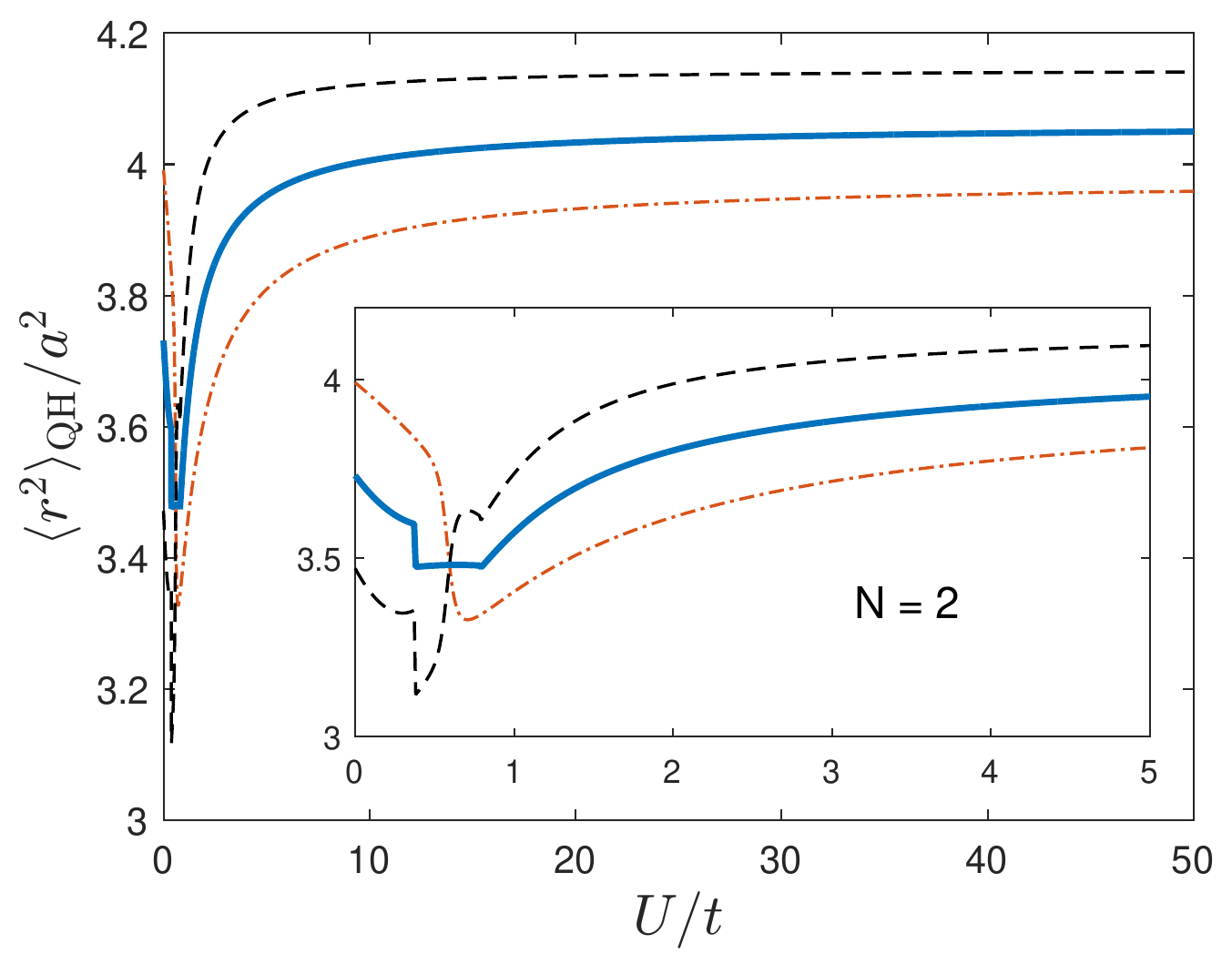}
\caption{Dependence of $\langle r^2\rangle_{\rm QH}$ on the interaction strength $U$ for $N=2$ obtained via exact diagonalization. Dash-dotted red line is calculated for the lowest-energy state, dashed black line is for the next-lowest-energy state, and solid blue line is the average of these two results. Inset shows the low-$U$ behavior.}
\label{SuppFig4}
\end{figure}

\begin{figure}[htbp]
\includegraphics[width=0.48\textwidth]{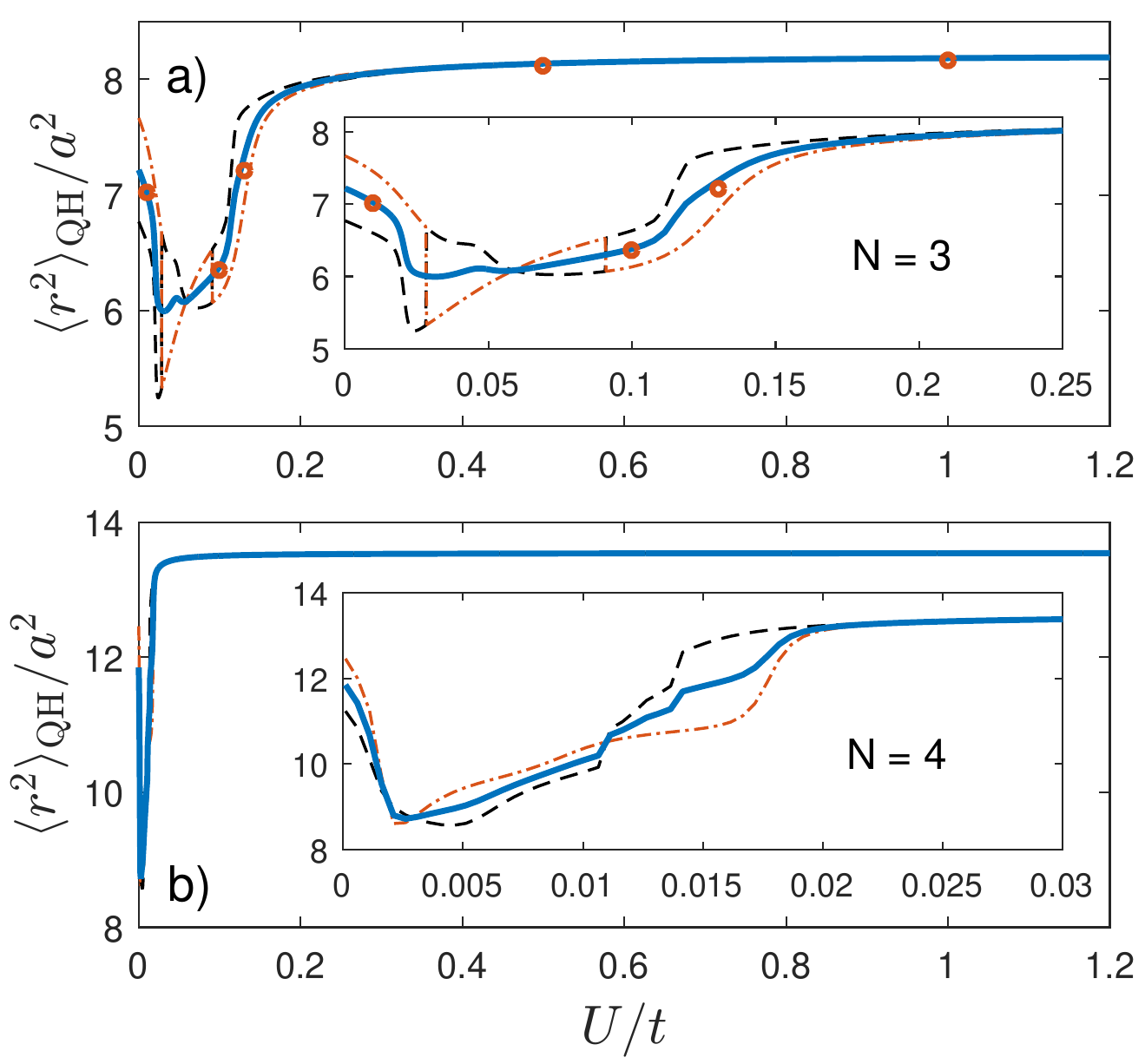}
\caption{Dependence of $\langle r^2\rangle_{\rm QH}$ on the interaction strength $U$ for (a) $N=3$ and (b) $N=4$ obtained in the lowest-band approximation. Dash-dotted red line is calculated for the lowest-energy state, dashed black line is for the next-lowest-energy state, and solid blue line is the average of these two results. Insets show the low-$U$ behavior. Also shown in (a) by red circles are the results for the average $\langle r^2\rangle_{\rm QH}$ calculated by exact diagonalization.}
\label{SuppFig5}
\end{figure}

In this part, we display how the interaction strength $U$ affects the $\langle r^2\rangle_{\rm QH}$ values. The general trend is that $\langle r^2\rangle_{\rm QH}$ evaluated for the two lowest-energy (nearly) degenerate states, although different for small $U$, quickly approach each other as $U$ is increased and converge to very close values for large $U$. However, the $N = 2$ case is an exception. In Fig. \ref{SuppFig4}, we show the results obtained via exact diagonalization for $N = 2$. Unlike the results for the systems with larger particle numbers displayed in Fig. \ref{SuppFig5}, $\langle r^2\rangle_{\rm QH}$ for the two lowest-energy states never do approach each other although they converge to certain values for large $U$. Even for large $U$ these states are only nearly degenerate. The reason for the discrepancy in $\langle r^2\rangle_{\rm QH}$ might then be attributed to the fact that the $N = 2$ system is a bit too small and slight differences in the densities get amplified in the lattice sampling of $\langle r^2 \rangle_{\rm QH}$. 

\section{Simple Disk Model}
\label{Disk model}

We show that the continuum expectation $\langle r^2\rangle_{\rm L}/\langle r^2\rangle_{\rm QH} = N/(N+1)$ can also be obtained from a simple model where the density profiles of the Laughlin and quasihole states can be taken as a disk and a disk with a hole, respectively. 

We suppose that the density of the incompressible Laughlin state is some constant $\rho$ up to a radius $R_{\rm L}$ and zero out of the disk. The total particle number is simply given by $N = \pi R_{\rm L}^2 \rho$. The mean square radius is then
\bea
\langle r^2 \rangle_{\rm L} = 2\pi \int_0^{R_{\rm L}}r^2\rho rdr/N = \frac{2\pi\rho}{N}\frac{R_{\rm L}^4}{4}=\frac{N}{2\pi\rho},
\label{r^2 L supp}
\eea
where $R_{\rm L}^2 = N/\pi\rho$ is used. Next, we punch a hole with radius $r_{\rm QH}$ at the center of the disk to model the quasihole. We assume that exactly one half of a particle is removed from this hole: $\pi r_{\rm QH}^2\rho = 1/2$. Supposing that the bulk density remains constant at the value $\rho$, the quasihole state will extend to a radius $R_{\rm QH}$ greater than $R_{\rm L}$. The particle number is still the same: $N = \pi(R_{\rm QH}^2-r_{\rm QH}^2)\rho$. Using this relation, we find
\bea
R_{\rm QH}^2-r_{\rm QH}^2 = \frac{N}{\pi\rho}\label{R^2-r^2},\\
R_{\rm QH}^2 = \frac{1}{\pi\rho}\bigg(N+\frac{1}{2}\bigg)\label{R^2},
\eea
where the second equation is obtained from $r_{\rm QH}^2 = 1/2\pi\rho$. The mean square radius for the quasihole state is found as
\begin{multline}
\langle r^2 \rangle_{\rm QH} = 2\pi \int_{r_{\rm QH}}^{R_{\rm QH}}r^2 \rho rdr/N \\
=\frac{2\pi\rho}{N}\frac{1}{4}\bigg(R_{\rm QH}^2-r_{\rm QH}^2\bigg)\bigg(R_{\rm QH}^2+r_{\rm QH}^2\bigg)\\
=\frac{1}{2}\bigg(R_{\rm QH}^2+r_{\rm QH}^2\bigg) = \frac{1}{2\pi\rho}(N+1),
\label{r^2 QH supp}
\end{multline}
where we used Eq. (\ref{R^2-r^2}) to obtain the first equality in the last line and Eq. (\ref{R^2}) together with the relation $r_{\rm QH}^2 = 1/2\pi\rho$ to obtain the last equality. Finally, dividing Eq. (\ref{r^2 L supp}) by Eq. (\ref{r^2 QH supp}) we arrive at the desired continuum ratio $N/(N+1)$. Subtracting Eq. (\ref{r^2 L supp}) from Eq. (\ref{r^2 QH supp}), one can also find an estimate for the quasihole radius as
\bea
r_{\rm QH} = \sqrt{\langle r^2 \rangle_{\rm QH}-\langle r^2 \rangle_{\rm L}}.
\label{r_QH}
\eea


\begin{thebibliography}{99}
\bibitem{note 2} As the many-body gap $\Delta$ above the ground state is a fraction of $U\phi$ \cite{Lukin FQH}, the typical values $\phi = 0.1$ and $U = t$ used in the calculations yield $\Delta \sim 0.1t$. The single-particle gap for $\phi = 0.1$ is $\sim 1.1t$, so the low-energy states are well described in the lowest-band approximation.
\bibitem{note} A symmetric lattice for the quasihole state with same $N$ can also be envisaged with $L_x^{\prime} = L_y^{\prime} = q+1$ and $\phi^{\prime} = 1/(q+1)$ to yield $N_{\phi}^\prime = q+1$. We numerically checked that such a configuration gives similar results.
\end{thebibliography}
\end{document}